\documentclass[10pt,letterpaper,twocolumn]{article}

\usepackage{ol2}
\usepackage[draft]{hyperref}
\usepackage{amsmath}
\usepackage{amssymb}

\begin{document}

\twocolumn[ 

\title{Mobility of high-power solitons in saturable nonlinear photonic lattices}

\author{Uta\ Naether and Rodrigo A.\ Vicencio}
\affiliation{Departamento de F\'isica, Facultad de Ciencias, Universidad
de Chile, Santiago, Chile, and Center for Optics and Photonics, Universidad de Concepci\'on, Casilla 4016, Concepci\'on, Chile}

\author{Milutin Stepi\'c}
\address{Vin\v ca Institute of Nuclear Sciences, University of Belgrade, P.O.B. 522, 11001 Belgrade, Serbia}

\begin{abstract}We theoretically study the properties of one-dimensional
nonlinear saturable photonic lattices exhibiting multiple mobility
windows for stationary solutions. The effective energy barrier
decreases to a minimum in those power regions where a new
intermediate stationary solution appears. As an application, we
investigate the dynamics of high-power gaussian-like beams finding
several regions where the light transport is enhanced.
\end{abstract}


 ]

\maketitle

For more than a decade, nonlinear waveguide arrays (WAs) have
become an excellent experimental scenario, where to study the
phenomenology appearing in periodic and non-periodic nonlinear
dynamical systems~\cite{rep1,rep2,rep3}. Different geometries,
dimensions, and nonlinearities have been studied showing
interesting and very different phenomenologies compared to
continuous systems. For WAs with Kerr nonlinearity, a coupled mode
approach leads to a discrete nonlinear Schr\"odinger (DNLS)
equation. For photorefractive media with saturable nonlinearity
the corresponding model is a s-DNLS~\cite{milutin}.

Mobility of localized solutions in nonlinear cubic WAs is well
known. As far as the power stays low, the energy barrier imposed
by the discreteness and the nonlinearity (usually called
Peierls-Nabarro (PN) potential~\cite{yuricam}) will stay small and
solutions will move across the lattice by just giving them a
judicious kick~\cite{switch1,switch2}. For larger powers, the
energy barrier grows and mobility is not possible anymore.
However, for saturable systems there are several regions of power
where the energy difference between the two fundamental localized
solutions - the one centered at one site (odd mode) and the one
centered between two sites (even mode) - vanishes for different
power values~\cite{milutin}. Close to these points, there are
regions of stability exchange between the even and odd solutions.
Since those regions exhibit bistability, the appearance of an
intermediate asymmetric and unstable solution is inevitable, as it
was shown in Ref.~\cite{2D,2Du} for two-dimensional systems.
Therefore, in this case, the \textit{effective} energy barrier
will strongly depend on the intermediate solutions (IS). For
saturable one-dimensional (1D) systems, this crucial issue has not
been clearly identified yet.We believe this element could be one of
the keys to experimentally observe, to the best of our knowledge
for the first time, good soliton mobility in 1D nonlinear saturable photonic lattices.

It was shown in Ref.~\cite{milutin} that the PN barrier becomes minimal exactly
at the points where the fundamental solution's energies coincide.
Contrary to what is expected, the fundamental solutions remain
immobile in these points. We demonstrate that in order to achieve
a good mobility it is necessary to increase the amount of power up
to the bifurcation point where the IS disappears. A constraint
method~\cite{2Du,1dsurface} is used to identify the ISs and
describe a pseudo-potential landscape among all stationary modes.
By using the system properties, we found recurrent resonant
behavior in power for gaussian-like shaped pulses showing enhanced
mobility.

In a 1D system the s-DNLS is given by
\begin{equation}
i \frac{d u_{n}}{d z}+(u_{n+1}+u_{n-1})-\gamma \frac{u_{n}}{1+|u_{n}|^2}=0\ ,\label{pde}
\end{equation}
where $u_n$ represents the light amplitude at site $n$, $\gamma$
the strength of the nonlinearity with respect to the coupling
coefficient, and $z$ a normalized propagation distance along the
waveguides. In order to understand the main phenomenology of these
lattices, we first look for stationary solutions of the form
$u_n(z)=u_n \exp(i\lambda z)$, where $u_n\in R$ and $\lambda$ is
the propagation constant or frequency. Small-amplitude plane waves define the
band $\lambda\in \left[-2-\gamma,2-\gamma\right ]$, while
high-amplitude plane waves define a second band $\lambda\in
\left[-2,2\right]$. Therefore in-phase stationary localized
solutions are limited to exist in the region $\lambda\in
\left[2-\gamma,2\right ]$, bifurcating from the fundamental modes
of those bands~\cite{2D}. Model (\ref{pde}) has  two dynamically
conserved quantities, the Hamiltonian
$H=-\left[\sum_{n}(u_{n+1}u^*_{n}+u_{n}u^*_{n+1})-\gamma
\log(1+|u_{n}|^2)\right ]$ and the optical power $P=\sum_n
|u_n|^2$.

\begin{figure}[h]
\centering
\includegraphics[width=0.45\textwidth]{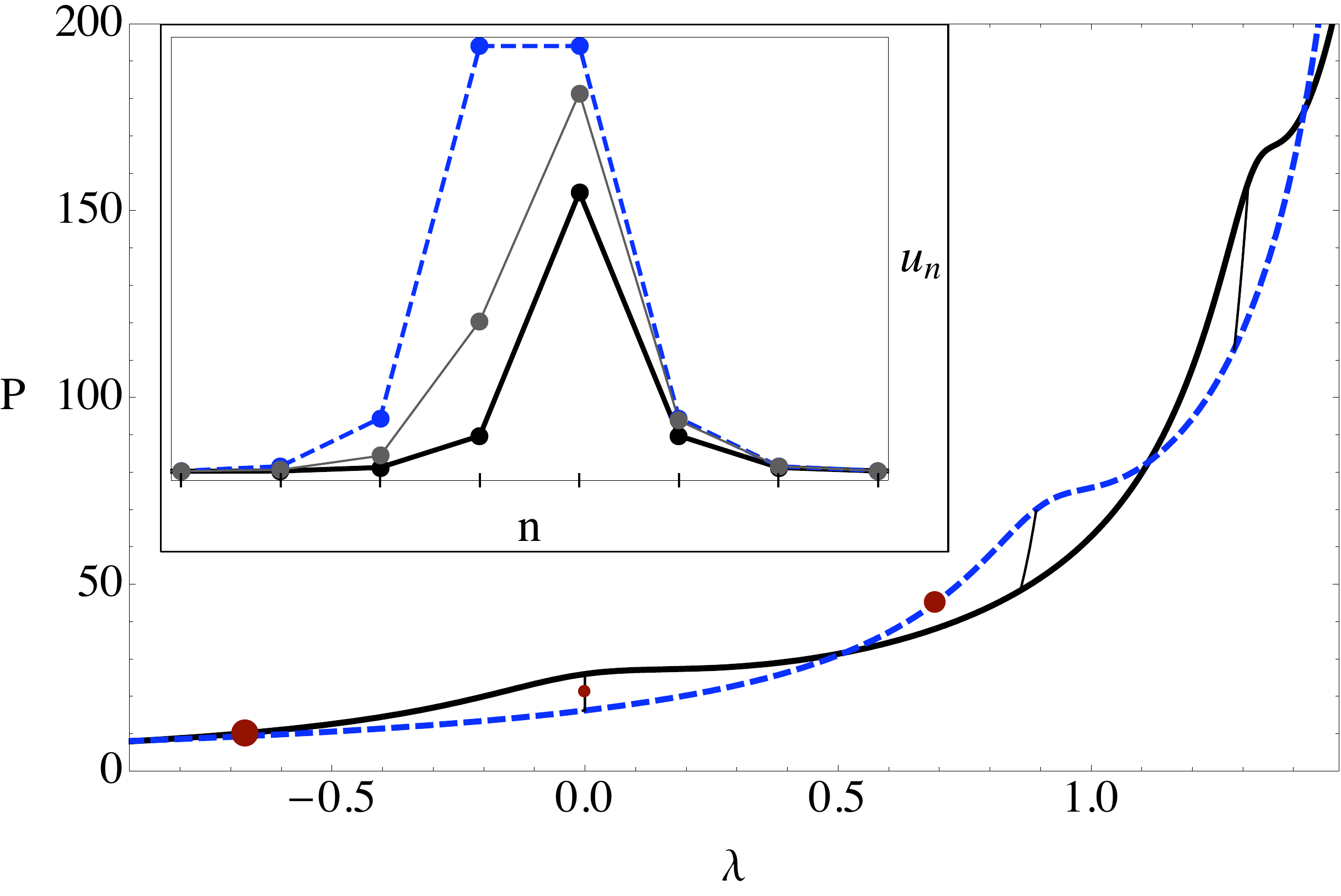}
\caption{(Color online) $P$ versus $\lambda$ for odd, even, and
intermediate solutions in full, dashed, and thin lines,
respectively. Inset: profiles corresponding to filled circles.}
\label{fig1}
\end{figure}
\begin{figure}
\centering
\includegraphics[width=0.45\textwidth]{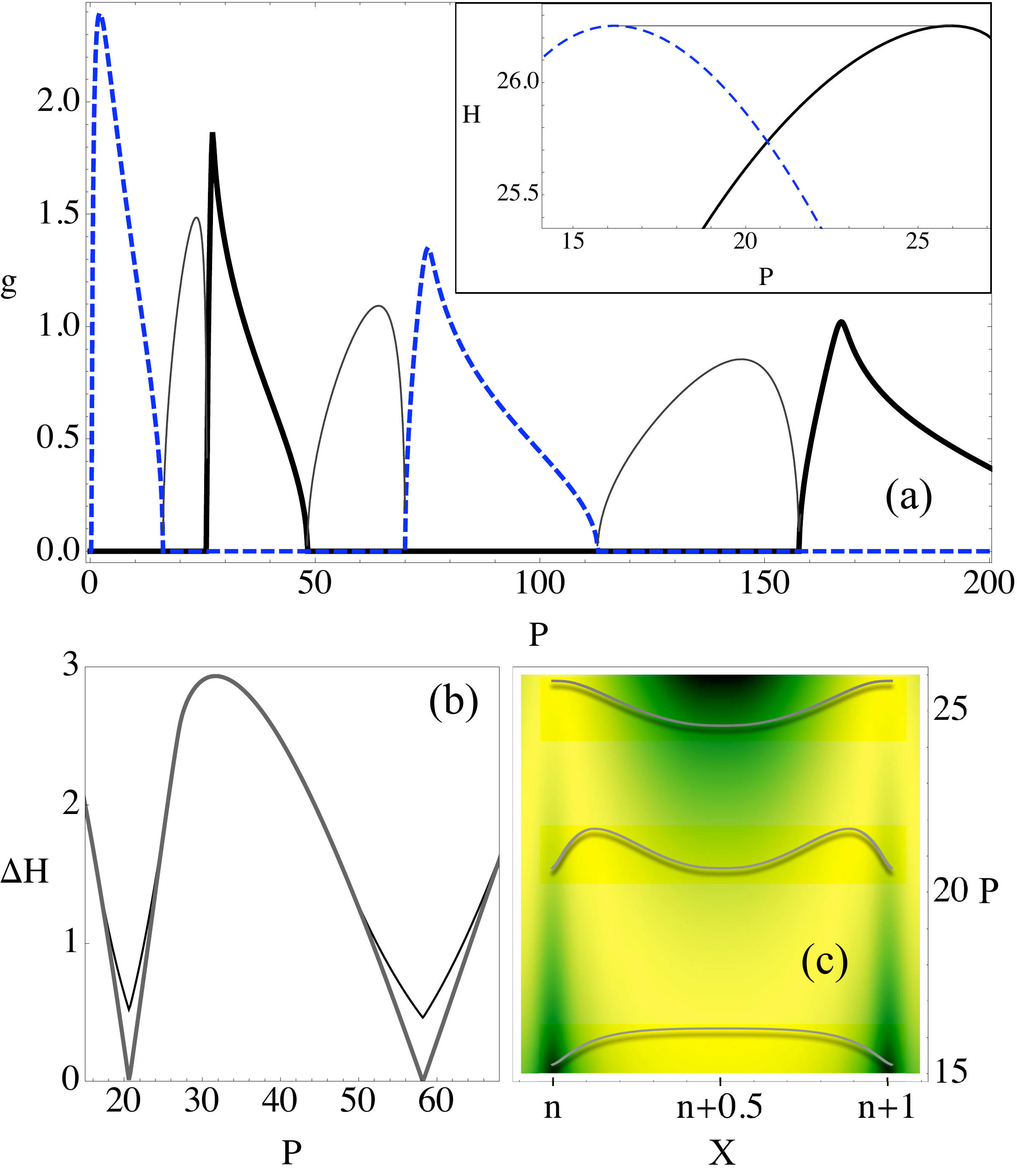}
\caption{(Color online) (a) $g$ versus $P$ for odd, even, and
intermediate solutions in full, dashed, and thin lines,
respectively. Inset: $H$ versus $P$. (b) $\Delta
H_0$ (thick line) and $\Delta H$ (thin line) versus $P$. (c)
Energy surface where light (dark) color denotes a high (low)
$H$-value.} \label{fig2}
\end{figure}

%
In the following, we discuss these properties for $\gamma=10$
(focusing nonlinearity). A different $\gamma$ will produce
different curves but the main saturable phenomenology will be
preserved. Localized solutions are computed by using a standard
Newton-Raphson method. Fig.\ref{fig1} shows a power versus
frequency diagram for both fundamental modes - the odd and the
even solutions - including the IS. As expected, the IS
corresponds to a non-symmetric profile connecting the two
fundamental modes [see Fig.\ref{fig1}-inset]. The two fundamental
solutions cross each other - repeatedly - as $P$ increases. In
regions where $|P_{odd}-P_{even}|$ is large, a family of IS
appears. It is remarkable that all the ISs of the first family
have $\lambda=0$ and connect both (even and odd) modes for this
value. In such a situation, the stationary solutions of Eq.(\ref{pde}) coincide with the ones of the
integrable Ablowitz-Ladik equation~\cite{prlmel}, which has an
analytic mobile solution. However, in the physical s-DNLS model,
it is not expected to find radiationless travelling solutions,
since mobile modes need to have the same power and not the same
frequency. We perform a standard linear stability
analysis~\cite{sta} by computing the largest imaginary part (``$g$'') of the eigenvalue spectrum. Fig.\ref{fig2}(a) shows our results
where $g=0$ implies stable solutions and $g>0$ unstable ones. For
all regions where the fundamental solutions are simultaneously
stable (three regions in this plot), the unstable IS appears. In
these regions there are points where the energy of both
fundamental solutions is exactly the same [see inset in
Fig.\ref{fig2}(a) for the first bistable region ($P\sim20$)].
However, the effective energy barrier is not zero if the IS is
considered as well. In Fig.\ref{fig2}(b) we plot $\Delta
H_{0}\equiv |H_{odd}-H_{even}|$ and $\Delta
H\equiv  |H_{max}-H_{min}|$ versus power. For the first two ``bistable'' regions, we clearly see that $\Delta H_{0}$ goes to
zero as it was previously predicted in Ref.~\cite{milutin}.
However, there is always a nonzero barrier ($\Delta H$) for the solution, which can be very small but it is - strictly speaking - nonzero. A first guess could be, that the most favorable region
for mobility would be the one where $\Delta H$ is a minima.
However, this is not the case for stationary solutions. If we kick
an odd or even mode, we are putting in motion an immobile-defined
solution, therefore there is always radiation from tails. As a
consequence, the power of the moving solution is lower than the
initial one. So, if we initially take the solution where $\Delta
H$ is a minima ($P_m$), the effective barrier will increase [see
Fig.\ref{fig2}(b)]. A better option would be the one where $P>P_m$
where, due to radiation, the power and the effective barrier
decrease. Now, in order to go deeper in the understanding of the
dynamics of 1D saturable WAs, we construct an energy landscape. By
defining the center of mass $X\equiv \sum_{n} n \vert
u_{n}\vert^2/P$ and using a constraint method\cite{2Du,1dsurface},
we compute $H$ versus $X$ and $P$. Fig.\ref{fig2}(c) shows our
computations around the first bistable region. In this plot, $X=n$
and $n+1$ correspond to odd modes while $X=n+0.5$ corresponds to
even one. For low power, the potential is cubic-like in the sense
that the odd mode is stable while the even one is unstable (lower
curve for $P=15$). By increasing the power, both fundamental
solutions are simultaneously stable because they are both a local
minima in this potential. Consequently, a maxima in-between
appears, the unstable IS (middle curve for $P=20.5$ where $\Delta
H_{0}\approx0$). Then, by further increasing the power, the odd solution
transforms into an unstable maxima while the even one becomes
the only minima of this potential (upper curve for $P=26$). The IS
originates when the even mode stabilizes; then it changes its
center of mass to an odd mode (symmetrically to the right and to
the left due to the system symmetry). Finally, the IS disappears
when the odd mode destabilizes.

%
%

\begin{figure}
\centering
\includegraphics[width=0.45\textwidth]{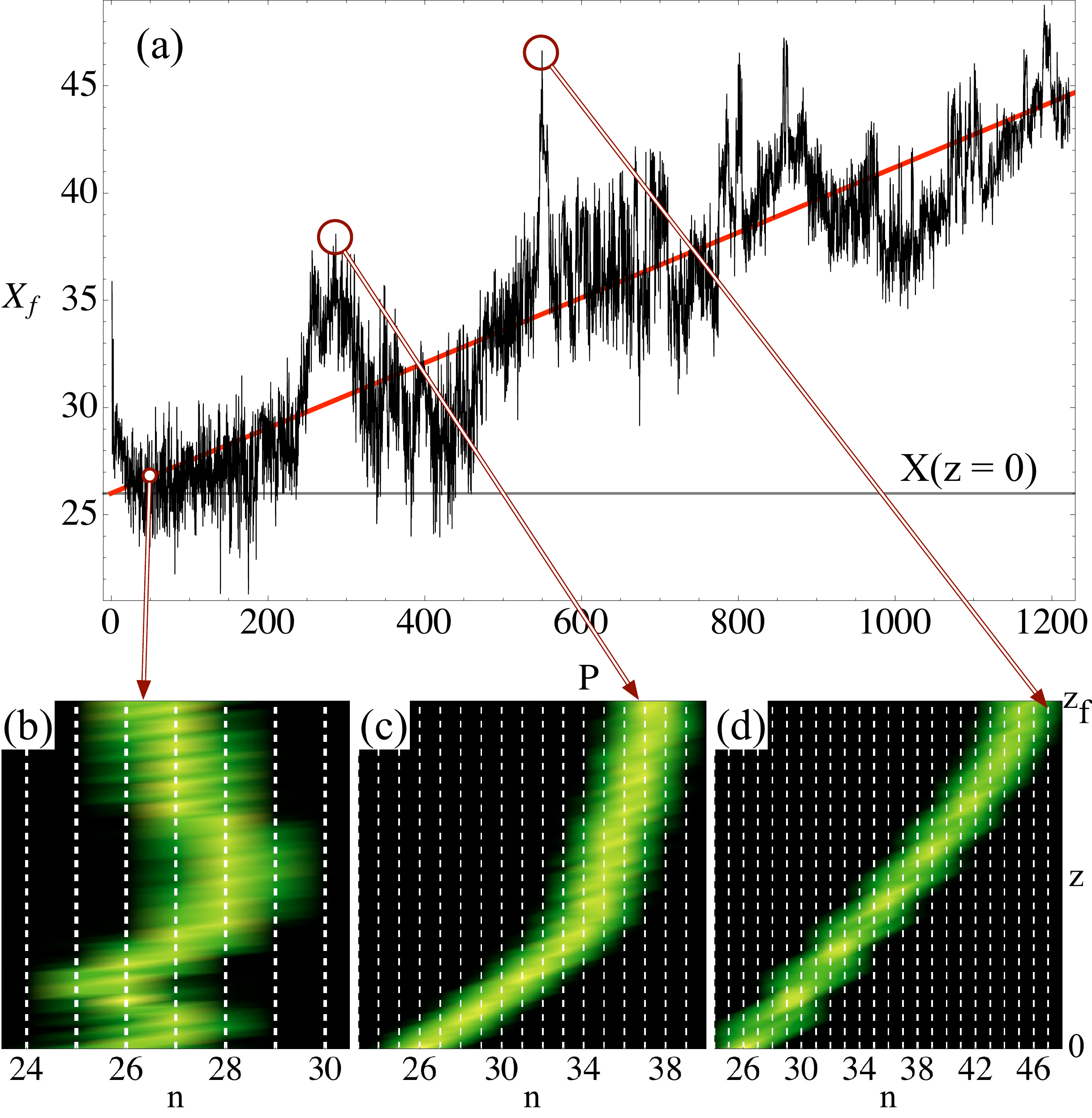}
\caption{(Color online) (a) Output $X_f$ versus input $P$. (b)-(d)
Dynamical examples for $P=54$, $300$, and $550$, respectively. $\gamma=10$,
$n_c=26$, $k=0.3$, $\alpha=1/3$, $z_{f}=50$.}
\label{fig3}
\end{figure}

To the best of our knowledge, mobility in this kind of systems was
never predicted for a more realistic experimental input condition
like gaussian input profiles. Previous simulations, starting from
stationary solutions, observed good
mobility~\cite{milutin,2D,2Du,prlmel}. However, saturable
solutions are not well localized in the power-exchange regions
and, furthermore, by increasing the power they become broader.
Therefore, in a experiment, dynamics will be strongly determined
by the power and shape of the chosen beam profile. We took as an
input beam a five-site wide gaussian-like profile: $u_n(0)=A
\exp[-\alpha(n-n_{c})^2] \exp[ik(n-n_{c})]$ for
$n=n_c,n_c\pm1,n_c\pm2$ and $u_n(0)=0$, otherwise. The kick $k$ is
proportional to the experimental angle and it does not alter the
power but adds a small amount of effective kinetic energy. With this initial condition, we numerically integrate model (\ref{pde}) from $z=0$ to $z=z_f$ and measure the center of mass at the lattice output: $X_f\equiv X(z_{f})$.
Fig.\ref{fig3}(a) shows our results for different input power ($P$). For very low power
($P\sim0$), the system is linear and mobility decreases as system
becomes nonlinear up to $P\approx 50$, when the saturable system
behaves as a cubic one [see Fig.\ref{fig3}(b)]. Then, by further
increasing the power, solutions start to move. When the power is
increased, fundamental solutions are geometrically similar and
differences in the Hamiltonian are very small, therefore the profile
is allowed to move with $k\neq 0$, as it is observed in the average
tendency of the curve [diagonal straight line in
Fig.\ref{fig3}(a)]. However, some ``resonant'' dynamics is found
for different levels of power. There are different regions where
mobility is enhanced as shown in Figs.\ref{fig3}(c) and (d), with almost undamped motion. It is
plausible to assume that this behavior corresponds to a
manifestation of the continuously repeating bistable regions
discussed for stationary solutions. Therefore, for even higher
powers, there should be good mobility, in principle without
limitations. The mobility windows for gaussian pulses do not
coincide with the mobility regions for stationary solutions, since
a gaussian pulse will always have a higher cost in power loss due
to radiation and due to its shape (gaussian profiles do not match in shape and power with any fixed stationary solution, what implies different level of powers to observe similar dynamics). But, nevertheless, the recurrent appearance of
enhanced mobility for certain powers shows an excellent phenomenological agreement
between gaussian and stationary profiles.

To conclude, we have shown that in nonlinear saturable 1D photonic lattices
there are several regions of bistability where stationary
solutions possess a small but nonzero energy barrier. The effective energy barrier
among all stationary localized solutions was constructed allowing
us to get a deeper understanding of discrete saturable nonlinear
systems. By using these properties with a more realistic input
condition, we were able to observe very good mobility and also to
find different regions of resonant response where the mobility is
enhanced. We hope that these findings will motivate
experimentalists to explore this direction of high-power mobility,
which is currently a drawback for the implementation of nonlinear WAs
in realistic all-optical operations in different power regimes.

Authors thank Magnus Johansson for discussions and acknowledge
financial support from: FONDECYT grants 1070897, Programa de
Financiamiento Basal de CONICYT (FB0824/2008), CONICYT fellowship,
and the Ministry of Science and Technological Development of
Republic Serbia (project III45010).

\end{document}